\documentclass[aps,prb,groupedaddress,reprint,showpacs]{revtex4-1}
\usepackage{graphicx}
\usepackage{tabularx}
\usepackage{dcolumn}
\usepackage{bm}
\usepackage{amsmath}
\usepackage{amssymb}
\usepackage{txfonts}
\usepackage[version=3]{mhchem}
\usepackage{url}
\usepackage[usenames,dvipsnames]{color}
\definecolor{Gray}{gray}{0.0}
\definecolor{lightGray}{gray}{0.35}

\begin{document}
\title{
  Electrochemical  Properties and Crystal Structure of \ce{Li+} / \ce{H+} Cation-exchanged \ce{LiNiO2}
}
\author{Takahiro Toma$^{1,2}$}
\author{Ryo Maezono$^{2,6}$}
\author{Kenta Hongo$^{3-6}$}

\affiliation{$^{1}$
  Battery Research Laboratories
  Sumitomo Metal Mining Co., Ltd., 
  Isoura-cho 17-3, Niihama, Ehime 792-0002, Japan
}

\affiliation{$^{2}$
  School of Information Science, JAIST, Asahidai 1-1, Nomi,
  Ishikawa 923-1292, Japan
}

\affiliation{$^{3}$
  Research Center for Advanced Computing Infrastructure,
  JAIST, Asahidai 1-1, Nomi, Ishikawa 923-1292, Japan
}

\affiliation{$^{4}$
  PRESTO, Japan Science and Technology Agency, 4-1-8 Honcho,
  Kawaguchi-shi, Saitama 322-0012, Japan
}

\affiliation{$^{5}$
  Center for Materials Research by Information Integration, 
  Research and Services Division of Materials Data and Integrated System, 
  National Institute for Materials Science, 1-2-1 Sengen, Tsukuba, Ibaraki 305-0047, Japan
}

\affiliation{$^{6}$
  Computational Engineering Applications Unit, RIKEN, 2-1 Hirosawa,
  Wako, Saitama 351-0198, Japan
}

\date{\today}
\begin{abstract}
\ce{LiNiO2} has high energy density but easily reacts  
with moisture in the atmosphere and deteriorates.
We performed qualitative and quantitative evaluations of the degraded phase of \ce{LiNiO2} and the influence of the structural change on the electrochemical properties of the phase.
\ce{Li_${1-x}$H_xNiO2} phase with cation exchange between \ce{Li+} and \ce{H+} was confirmed by 
thermogravimetric analysis and Karl Fischer titration measurement.
As the \ce{H} concentration in \ce{LiNiO2} increased, the rate capability deteriorated, 
especially in the low-temperature range and under low state of charge.
Experimental and density functional theory (DFT) calculation results suggested that this outcome was due to increased activation energy of \ce{Li+} diffusion owing to cation exchange.
Rietveld analysis of X-ray diffraction and DFT calculation confirmed that the $c$ lattice parameter and Li-O layer reduced because of the
\ce{Li+}/\ce{H+} cation exchange.
These results indicate that \ce{LiNiO2} modified in the atmosphere has a narrowed Li-O layer, 
which is the \ce{Li} diffusion path, and the rate characteristics are degraded. 
\end{abstract}
\maketitle
\section{Introduction}
Increasing environmental awareness has raised the demand for vehicles with low \ce{CO2} emissions.
In recent years, global environmental concerns have accelerated the development and spread of hybrid electric vehicles (HEVs), plug-in hybrid electric vehicles, pure electric vehicles, and fuel-cell-based electric vehicles equipped with lithium-ion batteries (LIBs).
LIBs installed in electric vehicles are required to have high energy density, rate capability, safety, and lifetime.~\cite{Neubauer2014}
Ni-rich cathode active materials such as \ce{LiNiO2}~\cite{Ohzuku1993,Kalyani2005,Moshtev1999,Ohzuku1995} 
and \ce{LiNi_{x}Co_{y}Al_{z}O2}~\cite{Lin2002,Zhu2004,Park2001,Li1997,Lee2001a,Liu2006} are used in
automotive lithium ion secondary battery applications owing to their high energy density.
However, Ni-rich cathode active materials have the problem that they easily react with moisture in the atmosphere,
causing structural change of the material and generation of residual Li impurities
such as \ce{LiOH}, \ce{Li2CO3}, and \ce{LiHCO3}.
As a result, the \ce{Li+} insertion and desorption reactions with the active material are inhibited, 
and the electrochemical characteristics are deteriorated.~\cite{Matsumoto1999,Liu2007,Faenza2017,Zhang2011,Liu2006}
Regarding the structural change caused by the reaction between \ce{LiNiO2} and \ce{H2O}, 
a structure in which \ce{Li} and \ce{H} are cation-exchanged as shown in eq (\ref{eq:LiHNiO2}) has been proposed.~\cite{Faenza2017,Moshtev1999}
However, a result that confirms the existence of \ce{Li_{1-x}H_xNiO2} experimentally has not been obtained.
\begin{equation}
\text{\ce{H2O} + \ce{LiNiO2} $\rightarrow$ $\left(1-x\right)$\ce{H2O} + x\ce{LiOH} + \ce{Li_{1-x}H_xNiO2}}
\label{eq:LiHNiO2}
\end{equation}
One of the reasons why it is difficult to identify substances experimentally is the extremely low content of the target.
Usually, X-ray powder diffraction is effective in identifying the generated heterogeneous phase, 
but if the amount of phase is very small, such as a phase that has been altered by exposure to the atmosphere, 
it may be impossible to evaluate it directly because it is below the detection limit of the measuring device.

\vspace{2mm}
Herein,  we performed experimental and computational approaches, 
to find out the qualitative and quantitative identification of the structural phase of \ce{Li_${1-x}$H_xNiO2}
and the investigation of its influence on the electrochemical properties.
The qualitative and quantitative identifications of structural change phases were attempted 
by combining thermogravimetric analysis and Karl Fischer titration measurements.
Although it may not be possible to clearly deduce the structural change phase itself in experiments, 
its influence on the electrochemical properties and lattice parameter can be clearly observed even with a small amount.
Therefore, we simulated the crystal structure and electrochemical properties of \ce{Li_${1-x}$H_xNiO2} 
which was estimated structural change phase, using first-principles calculations and confirmed the agreement between the experimental and simulation results.
This study is a new initiative that shows the effects of atmospheric exposure of cathode active materials on the crystal structure and electrochemical properties accurately and quantitatively, and proposes factor to be considered especially for Ni-rich cathode active materials.
\label{section.introduction}
\section{Experimental methods}
\subsection{Substitution of \ce{Li+}/\ce{H+} in \ce{LiNiO2}}
\ce{LiNiO2} samples were obtained by solid phase synthesis of \ce{Ni(OH)2} and \ce{LiOH}.
\ce{Ni(OH)2} precursors were synthesized by co-precipitation.~\cite{Huang2015a}
A 2.0 M aqueous solution of \ce{NiSO4} was pumped into a continuously stirred tank reactor 
in nitrogen atmosphere.
The \ce{pH} of the solution inside the reactor was controlled at approximately 11 by adding \ce{NaOH} solution. 
\ce{NH4OH} solution, which acts as a chelating agent to control the \ce{Ni} ion neutralization reaction time,
was separately pumped into the reactor.
The \ce{Ni(OH)2} powder sample was washed, filtered, and dried in an oven at 120 $^\circ$C for several hours.
The prepared precursors were thoroughly mixed with \ce{LiOH} powder at the ratio of \ce{Li}/\ce{Ni} = 1.01.
The mixture was synthesized for 12 h at 710 $^\circ$C in \ce{O2} flow.
The synthesized \ce{LiNiO2} was placed in a chamber (LHL-113, ESPEC) maintained
at 30 $^\circ$C and 60\% relative humidity under air circulation for 0, 9, 25, and 49 h. The four prepared \ce{LiNiO2} samples are denoted as LNO$\_$0 h, LNO$\_$9 h, LNO$\_$25 h, and LNO$\_$49 h depending upon the exposure time.

\vspace{2mm}
\ce{LiNiO2}  contains \ce{LiOH} and \ce{Li2CO3} produced by exposure to the atmosphere;  
therefore, the effects of the structural change phase and the residual Li compounds cannot be distinguished.
In order to isolate the influence of the structural change phase, 
we attempted to remove the residual \ce{Li} compound by washing the active material.
The solvent used for removing the residual \ce{Li} compound should be selected by considering its reactivity with \ce{LiNiO2}.
For example, excessive extraction of \ce{Li} (eq.(~\ref{eq:LiHNiO2})) 
from the bulk can occur by using water as the solvent.
Excessive delithiation causes significant degradation of the electrochemical properties, ~\cite{Moshtev1999,Xu2017a}
making it difficult to evaluate the effects of atmospheric exposure properly.
Therefore, we used ethanol, which has poor reactivity with \ce{Li}, as the washing solvent.
To obtain \ce{LiNiO2} without residual \ce{Li} compound, 10 g \ce{LiNiO2} powder was added to 200 mL of distilled ethanol 
under constant stirring with a magnetic stirrer for 5 min.

\subsection{Material characterization}
 The surface morphologies of the samples were observed by field emission scanning electron microscopy (FE-SEM, JSM-7001, JEOL).
The phase transformation of the surface layer was observed
by spherical-aberration-corrected scanning transmission electron microscopy (CS-STEM, JEM-ARM200F, JEOL).
The samples were sliced to a thickness of 100 nm or less using a focused ion beam (FIB, NIB-5000, Hitachi),
and the samples were observed under the acceleration voltage of 200 kV.
The crystalline phases of the prepared samples were identified using X-ray powder diffractometer  
(XRD,  X'Pert Pro MPD, PANalytical) with \ce{Cu}-K$\alpha$ radiation.
The XRD data pattern was recorded at the scan rate of 5$^\circ$ min$^{-1}$
in the 2$\theta$ range of 15$^\circ$ to 100$^\circ$.
Rietan-FP software~\cite{Izumi2007} was used for Rietveld analysis.
The thermal behavior of the \ce{LiNiO2} samples were investigated by thermogravimetric analysis (TGA) (DSCX3100SA, NETSZCH).
Karl Fischer (KF) titration (CA-200, MITSUBISHI CHEMICAL ANALYTICAL) was used to evaluate to \ce{H2O} concentration.

\subsection{Electrochemical techniques}
 Electrochemical measurements were carried out with a laminated-type full cell and CR2032 coin-type symmetric cell with positive
electrodes facing each other. These cells were used for rate capability test and
electrochemical impedance spectroscopy (EIS) measurement, respectively.
Each \ce{LiNiO2} sample was mixed with a conducting agent (carbon black) and polyvinylidene fluoride (PVDF) at 85:10:5 weight ratio
in N-methyl-2-pyrrolidone (NMP) to prepared a slurry.
Then, each slurry was applied to an aluminum current collector and dried in a vacuum oven for 8 h at 120 $^\circ$C.
The dried coating foil was then compressed to a layer of 20 $\mu$m thickness using a roll press.
Cu foil coated with graphite was used as the negative electrode in the laminate type cell.
The electrolyte was 1.0 M \ce{LiPF6} in ethylene carbonate (EC) - dimethyl carbonate (DMC) solution (3:7 by volume).
For the rate capability test of the laminate-type cells, discharge rates at different current densities varying from 0.1 to 10C were applied using
constant current (CC) with fixed charge rate of 0.1C and constant current - constant voltage (CC-CV) within the voltage range from 2.5 to 4.2 V at 25 $^\circ$C.   
EIS measurements of the symmetric cells were carried out by applying an AC - amplitude of 10 mV at 100 kHz to 1 mHz frequency
between -20 $^\circ$C and 45 $^\circ$C. The resulting Nyquist plots were analyzed using ZView software (Scribner Associates, Inc.).

\subsection{Computational Methodology}
All first-principles calculations using DFT were performed with
the Vienna Ab initio Simulation Package (VASP)~\cite{Kresse1996,Kresse1996a}
with projector augmented wave (PAW) potentials.
A plane-wave basis set cutoff energy of 650 eV and a Monkhorst-Pack 3 × 3 × 1 k-points mesh was used for all calculations.
The calculation was performed using a supercell constructed from
the primitive cell by tripling the system along the directions of the $a$ and $b$ axes.
Energy optimizations were based on the tetrahedron method with Bl\"{o}chl corrections for Brillouin-zone integrations;
their convergence criteria for the self-consistent field (SCF) and maximum force were set to 0.1 meV/atm and 1.0 meV/\AA, respectively.
We adopted the optB86b-vdW functional~\cite{Klime2011,Yoshida2019,Radin2016} as the exchange-correlation (XC) functional
because it is known to accurately reproduce experimental results such as voltage and volume in the Li detachment state.
Considering the Coulomb interaction of the Ni 3$d$ orbit, a correction of $U = 6.7 eV$ was added to Ni 3$d$.~\cite{Zhou2004a}
In this study, the structural configuration was drawn using VESTA software.~\cite{Momma2011}
In the structure relaxation calculation of the Li desorption structure, all symmetry structures 
that are Li defect sites were calculated, and the lowest energy configuration was defined as the Li desorption structure.
The climbing image nudged elastic band (CI-NEB) method was used to determine the Li-ion diffusion barriers and paths.~\cite{Henkelman2000,Henkelman2000a}
The initial and final configurations were determined by separately optimizing
the structure with a vacancy at each of the two adjacent sites.
A group of images generated by linear interpolation between the two endpoint configurations was
optimized to converge to points on the minimum energy path.
The five intermediate images between the two endpoints were generated 
by linear interpolation between the two endpoint configurations.
The convergence criterion used in the CI-NEB calculations was that
the forces acting on each atom was lesser than 0.02 meV/\AA.
\label{section.experiment}
\section{Results and Discussion}
\subsection{Characterization}
\begin{figure*}
  \begin{center}
   \includegraphics[width=\hsize]{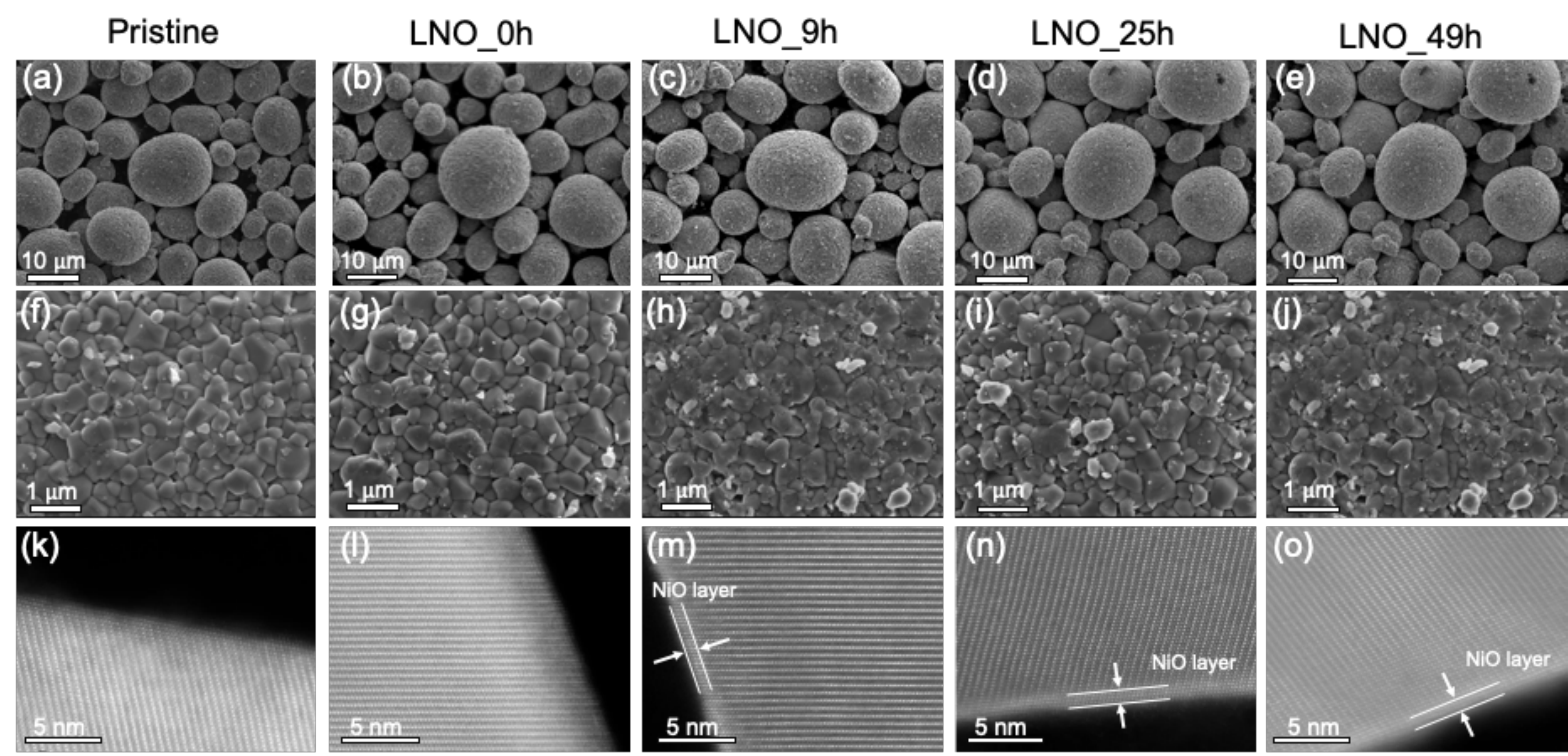}
   \caption{Low-magnification surface SEM images (a)-(e) and high-magnification SEM images (f)-(j) of the samples.
			Cross-sectional HAADF-STEM images ((k)-(o)).
			(a),(f),(k) Pristine samples (no exposure to air, no washing); (b),(g),(l) LNO$\_$0 h;
			(c),(h),(m) LNO$\_$9 h;
			(d),(i),(n) LNO$\_$25 h; and (e),(j),(o) LNO$\_$49 h panels.}
   \label{fig.fig1}
  \end{center}
 \end{figure*}

SEM and STEM observations were performed to evaluate the changes in the primary and secondary particles and surface structure changes 
before and after exposure to the atmosphere and washing.
Each sample was confirmed to be free of primary and secondary particle size changes, particle collapse, and deformation (Figure \ref{fig.fig1}(a)-(j)).
On the other hand, the particle surface changed smoothly as the exposure time to the atmosphere increased, 
suggesting structural changes due to exposure.
In the HAADF-STEM images of the particle surface (Figure \ref{fig.fig1}(k)-(l)),  
layered stripes representing the R\=3{m} structure of \ce{LiNiO2} were observed both inside and on the surface of pristine (k) and LNO$\_$0 h(l).
The results also confirmed that the washing process in this study removed the residual Li compound
without changing the structure of the active material. 
In contrast, the LNO$\_$9 h(m), LNO$\_$25 h(n), and LNO$\_$49 h(o) samples had altered layers 
with thickness of approximately 1 nm on the outermost surface. 
This alteration phase on the outermost surface was suggested to be \ce{NiO} with Fm\={3}m structure 
caused by the change in the arrangement of Ni atoms.~\cite{Lee2018,Cho2013,Hayashi2014,Makimura2012,Huang2019}
Because the \ce{NiO} thickness was almost constant regardless of the samples exposed to the atmosphere, 
the influence of \ce{NiO} is considered not to appear as a difference between the samples in the results of the electrochemical evaluation described later.
 
\begin{figure}
 \includegraphics[width=3.33in]{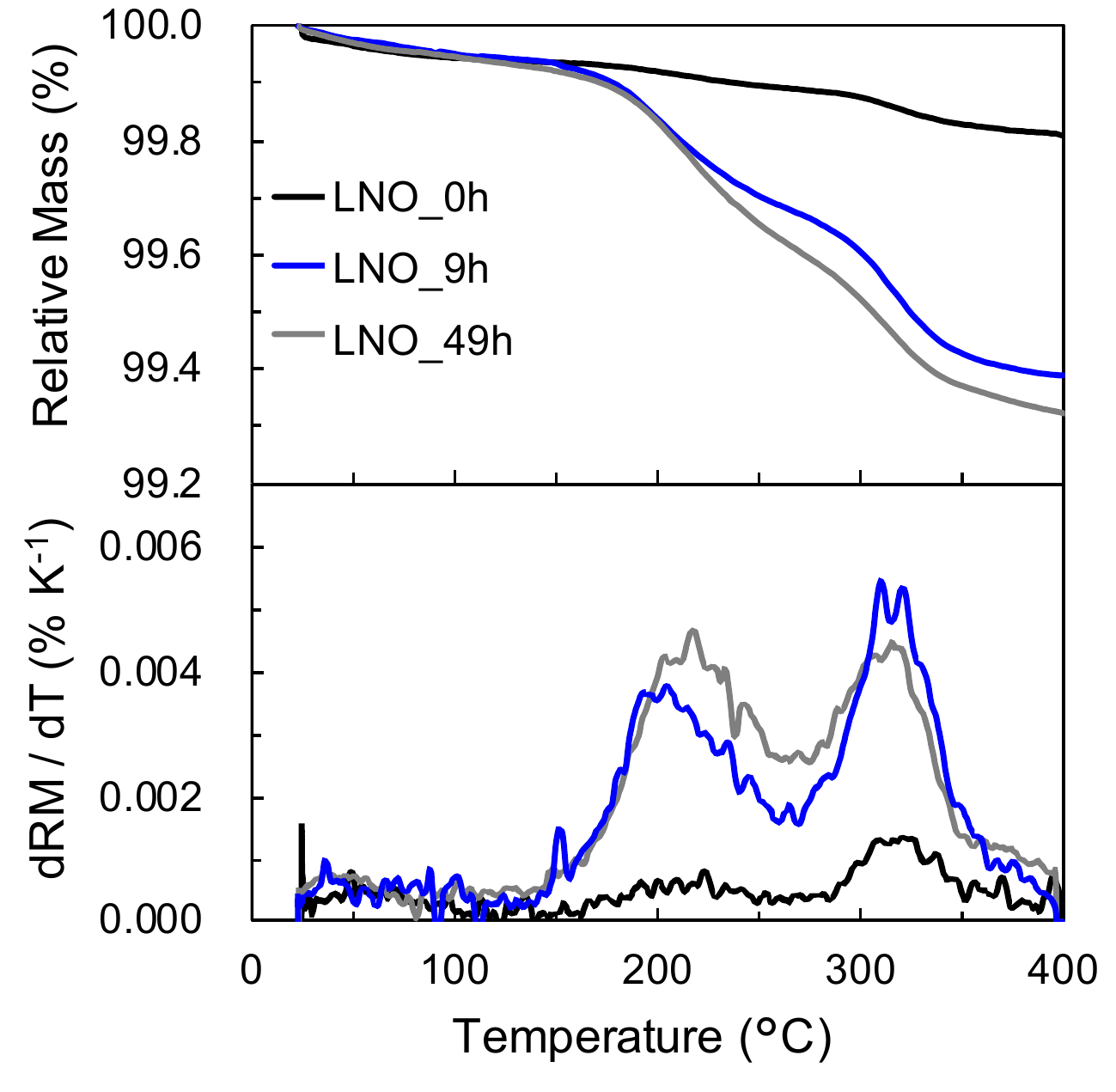}
 \caption{Normalized thermogravimetric rate of change of \ce{LiNiO2} exposed to air for 0 h, 9 h, and 49 h.
               The upper panel shows the results of heating from 25 $^\circ$C to 400 $^\circ$C in Ar atmosphere.
               The lower panel shows the result of differentiating the weight change rate with temperature (dRM/ d$T$).}
 \label{fig.fig2}
\end{figure}

TGA analysis was performed to identify the structural change phases owing to phenomena such as thermal decomposition and oxidation.
Figure \ref{fig.fig2} shows the TGA results of \ce{LiNiO2} after atmospheric exposure.

Weight loss from approximately 100-120 $^\circ$C indicates water adsorption--desorption reactions.
In the higher temperature region, large weight loss peaks were observed at 200--220 $^\circ$C and 300--320 $^\circ$C.
These weight change peaks have been reported to be due to \ce{H2O} and \ce{O2} elimination 
from \ce{NiOOH}, as shown in eq. (\ref{eq:NiOOH to NiO}).~\cite{Gu2011,Pan2005}
\ce{Ni(OH)2} and \ce{NiOOH} may also be generated by the mixing of \ce{Li+}/\ce{H+}.
However, the \ce{Ni(OH)2} decomposition temperature is approximately 300$^\circ$C, 
and the existence of one weight loss peak has been reported. ~\cite{Tang2014,Zhu2011}
Therefore, the phase generated by atmospheric exposure is considered to be \ce{NiOOH}. 
 \begin{equation}
 \text{\ce{NiOOH} $\rightarrow$ \ce{NiO} + $\frac{1}{4} \ce{O2}$ + $\frac{1}{2}$ \ce{H2O}}
 \label{eq:NiOOH to NiO}
 \end{equation}
Because LiNO$\_$0h had little weight reduction reaction in the temperature ranges of 200--220 $^\circ$C and 300--320 $^\circ$C, 
the amount of \ce{NiOOH} in the active material is considered to be small.
On the other hand, in LNO$\_$9h and LNO$\_$49h, two peak changes at 200 $^\circ$C  or higher increased 
with increasing exposure time.
From these results, it was suggested that the cation exchange of \ce{Li+}/\ce{H+} proceeded at atmospheric exposure,
and the production of \ce{NiOOH} increased.

\begin{figure}
 \includegraphics[width=3.33in]{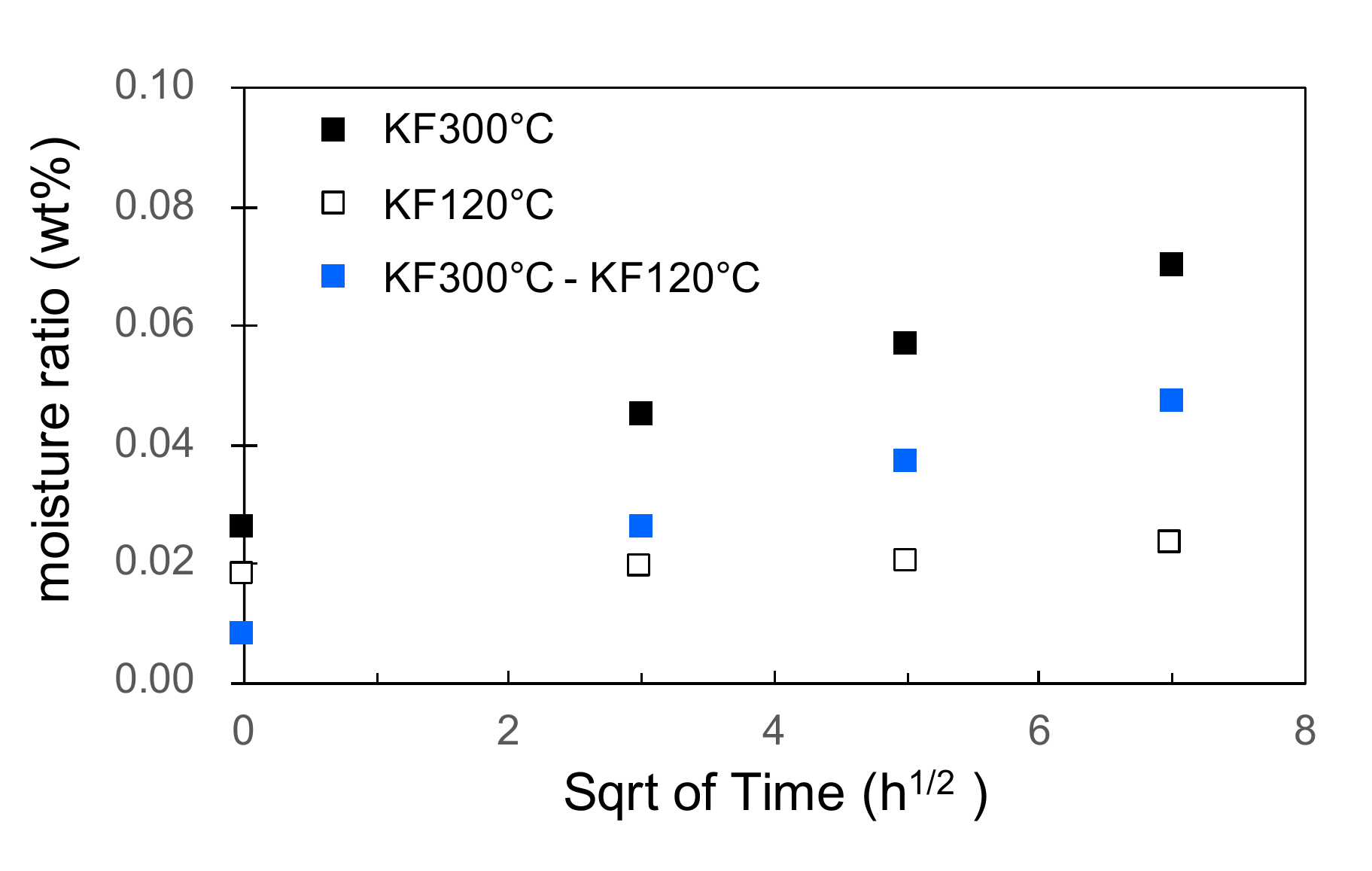}
 \caption{Relationship between atmospheric exposure time and moisture content of \ce{LiNiO2}.
              The black square and the white square are the titration results at 300 $^\circ$C  and 120 $^\circ$C, respectively.
              The blue square shows the difference between the results at 300 $^\circ$C  and 120 $^\circ$C.}
 \label{fig.fig3}
\end{figure}

TGA can estimate the relationship between temperature and weight change reaction,
but it is difficult to identify and quantitatively analyze the reacted substance.
Therefore, quantitative analysis of the amount of \ce{NiOOH}, 
that is, the amount of \ce{H+} exchange, was attempted 
by quantifying the amount of \ce{H2O} produced by the reaction in eq.(\ref{eq:NiOOH to NiO}).
In order to separate the amount of \ce{H2O} derived from the adsorbed water 
and the amount of \ce{H2O} derived from the decomposition reaction of \ce{NiOOH}, 
KF titration analysis was performed under two conditions of 120 $^\circ$C and 300 $^\circ$C.
Titration at 120 $^\circ$C detects the adsorbed water, 
and that at 300 $^\circ$C detects the total amount of \ce{H2O} generated by decomposition of adsorbed water and \ce{NiOOH}.
Therefore, the amount of \ce{H2O} derived from the decomposition of \ce{NiOOH} can be determined 
by taking the difference between the titration results of 300 $^\circ$C (KF 300$^\circ$C) and 120 $^\circ$C(KF 120$^\circ$C).
Figure \ref{fig.fig3} shows the results of KF titration.
The value of KF300$^\circ$C-KF120$^\circ$C increased in proportion to the square root of the exposure time,
that is, the amount of \ce{NiOOH} produced increased with the exposure time.
We assumed that the \ce{H} desorption reaction was due to heating during KF titration, as expressed in eq.(\ref{eq:KF proton}).
In this case, the amount of \ce{H} contained in \ce{LiNiO2} can be calculated as shown in eq.(\ref{eq:H conc}) by 
using the amount of \ce{H2O} derived from NiOOH, $\alpha_{\ce{H2O}}$ ($\Delta$ KF 300$^\circ$C - KF 120$^\circ$C).
The \ce{H} concentrations of LNO$\_$0 h, LNO$\_$9 h, LNO$\_$25 h, and LNO$\_$49 h were determined to be
0.08, 0.28, 0.40, and 0.51 at\%, respectively.
\begin{equation}
        \text{\ce{Li_{1-$\delta$}H_{$\delta$}NiO2} + $\frac{\delta}{4}$ \ce{O2} $\rightarrow$ \ce{Li_{1-$\delta$}NiO2} + $\frac{\delta}{2}$ \ce{H2O}}
     \label{eq:KF proton}
     \end{equation}
     \begin{equation}
     \text{$\frac{\delta}{2}$ = $\frac{\alpha_{\ce{H2O}} \cdot {M_{\ce{LiNiO2}}}}{{M_{\ce{H2O}}}}$}
     \label{eq:H conc}
   \end{equation}

\begin{figure}
   \includegraphics[width=3.33in]{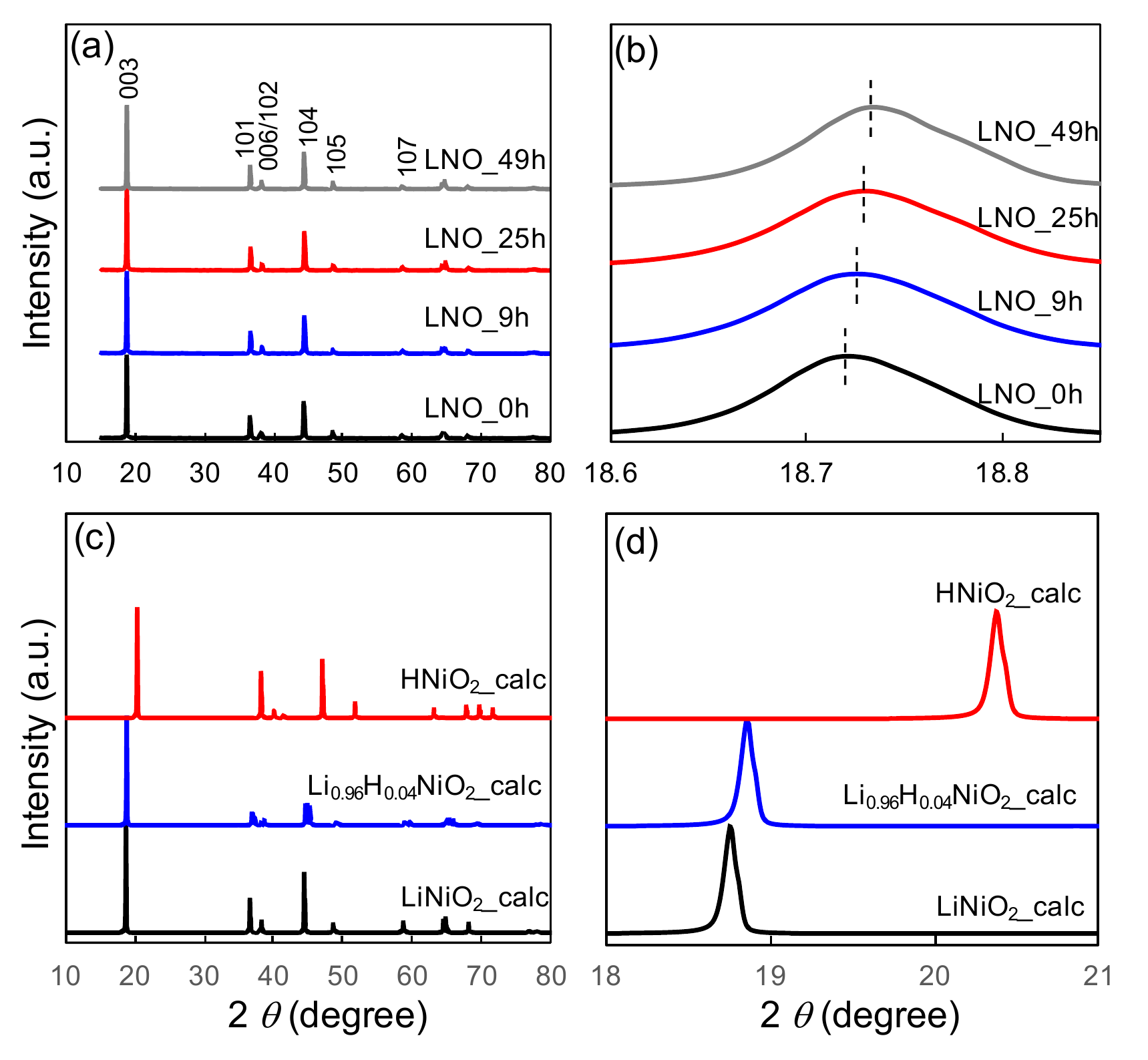}
   \caption{Panel (a) shows the entire XRD profile of each sample, 
            and panel (b) shows an enlarged view near the (003) diffraction angle.
            Panel (c) shows the result of simulating the XRD profile of \ce{LiNiO2} with 0, 3.7,
            and 100 at\% substitution of \ce{Li} with \ce{H} 
            by DFT calculation, and panel (d) is an enlarged graph of (003) diffraction angle of profile (c).}
   \label{fig.fig4}
  \end{figure}

XRD evaluation was performed to investigate the crystal structure change due to cation exchange of \ce{Li+}/\ce{H+}.  
Figure \ref{fig.fig4}(a),(b) shows the XRD profile of each sample.
From the results of Fig. \ref{fig.fig4}(a), it was confirmed that no significant heterogeneous formation was observed in any sample,
and \ce{LiNiO2} was the main phase.
The (003) diffraction peak near 18.8$^\circ$, i.e., the 3a site (Li site) in the R\=3{m} structure, 
showed a shift of the diffraction angle toward the higher angle side with 
increasing atmospheric exposure time (Figure \ref{fig.fig4}(b)).
In order to verify whether this peak shift was caused by \ce{Li+}/\ce{H+} cation exchange, 
the XRD peak of \ce{Li_{1-x}H_xNiO2} was simulated using first principles calculations.
Figure \ref{fig.fig4}(c), (d) shows the XRD profiles calculated
for the crystal structure of \ce{LiNiO2} obtained 
by structural relaxation calculation with Li replaced by 0 at\%, 3.7 at\%, and 100 at\% of \ce{H}.
Similar to the experimental results, the relationship between the \ce{H} content 
and the shift of the (003) peak position was also confirmed by calculation.
From these results, it was found that the (003) peak position shifts to the higher angle side due to the exchange of \ce{Li+}/\ce{H+}, i.e., the $c$-axis lattice constant shrinks.
In addition, the lattice constants obtained by Rietveld analysis and first-principles calculation are summarized in Table \ref{table1}.
\begin{table} 
 \caption{Lattice constant of \ce{Li+}/\ce{H+} cation exchange structure obtained by Rietveld analysis and first-principles calculation}
 \begin{tabular}{lccc} 
   \hline
   Samples & H (at\%) &a (\AA) & c (\AA) \\
   \hline
   LNO$\_$0h & 0.08 & 2.8776 & 14.2057 \\
   LNO$\_$9h & 0.28 & 2.8766 & 14.2005 \\
   LNO$\_$25h & 0.40 & 2.8761 & 14.1988 \\
   LNO$\_$49h & 0.51 & 2.8755 & 14.1962 \\
   \ce{LiNiO2}(calc) & 0 & 2.8500 & 14.1966 \\
   \ce{Li_{0.96}H_{0.04}NiO2}(calc) & 3.70 & 2.8446 & 14.1052 \\
   \ce{HNiO2}(calc) & 100 & 2.7554 & 12.9179 \\
   \hline
  \end{tabular}
 \label{table1}
\end{table}

\subsection{Electrochemical properties}
\begin{figure}
  \includegraphics[width=3.33in]{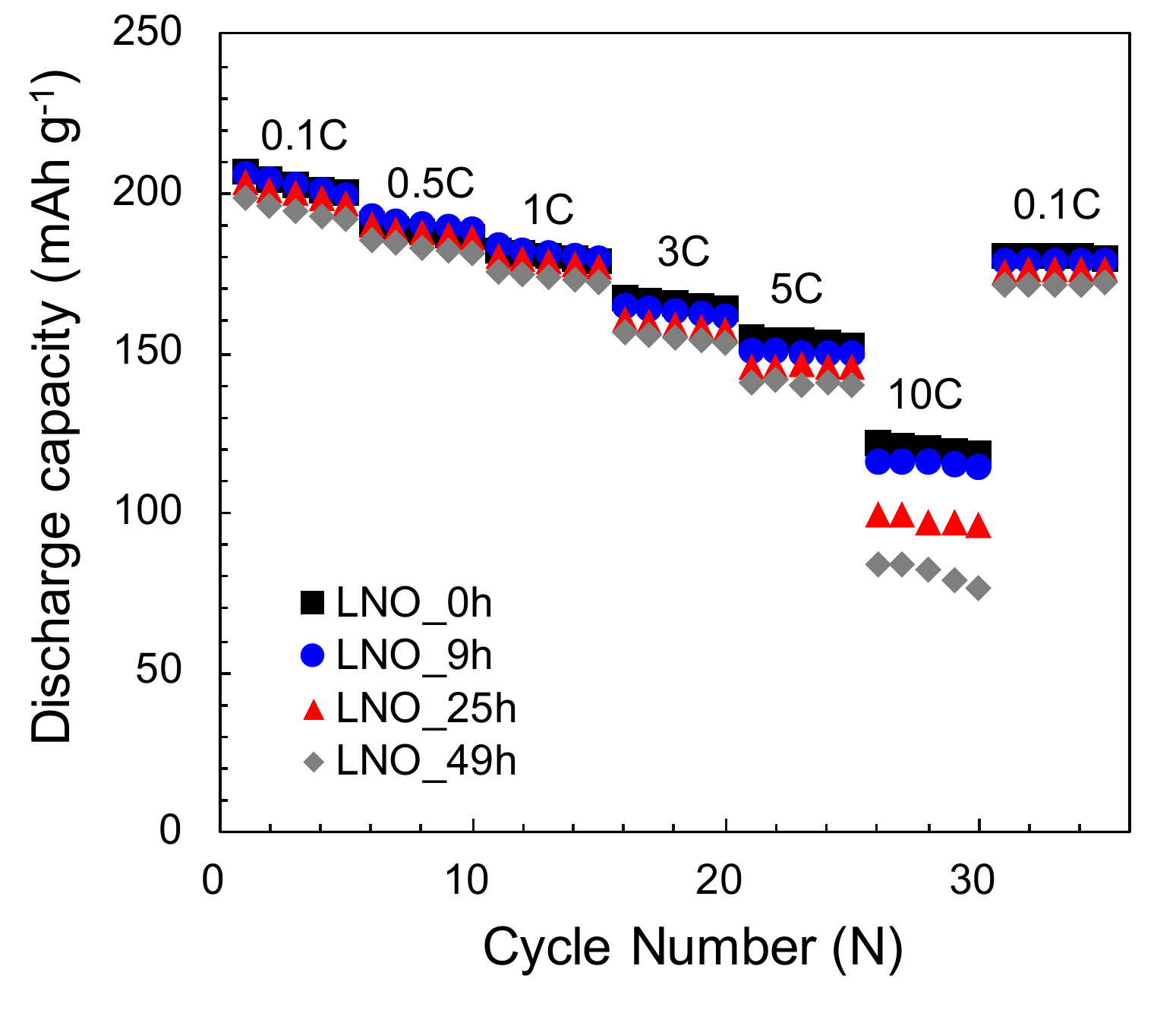}
  \caption{Rate characteristics measured at different rates of 0.1--10C. 
               The samples are denoted as black square (LNO$\_$0 h), blue circle (LNO$\_$9 h), red triangle (LNO$\_$25 h),
              and gray rhombus(LNO$\_$49 h). }
 \label{fig.fig5}
\end{figure}

Rate capability test results measured at different current densities (0.1--10C) are shown in Figure \ref{fig.fig5}.
At low  discharge rate (0.1--1C), all samples showed almost the same discharge capacity.
On the other hand, at high discharge rate (3--10C), a significant decrease in the discharge capacity was confirmed 
in the LNO$\_$25 h and LNO$\_$49 h samples.
This result implies that a material with a large amount of \ce{Li+}/\ce{H+} is slow to exhibit the
\ce{Li+} transfer reaction at high discharge rate.

\begin{figure}
 \includegraphics[width=3.33in]{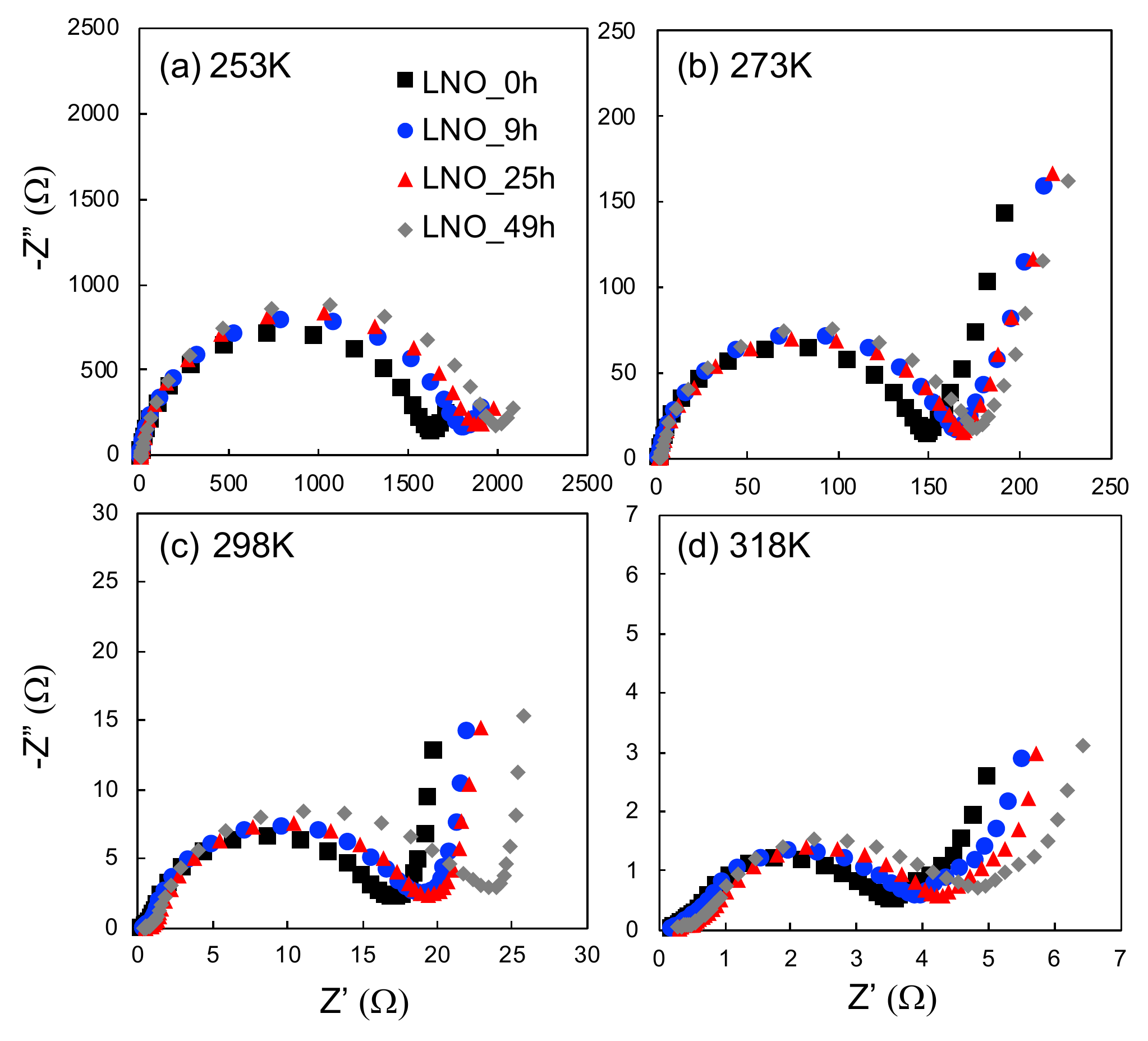}
 \caption{EIS spectrum of \ce{Li$_{0.50}$NiO2} at (a) 253 K, (b) 273 K, (c) 298 K, (d) 318 K}
 \label{fig.fig6}
\end{figure}

\begin{figure}[tb]
 \includegraphics[width=3.33in]{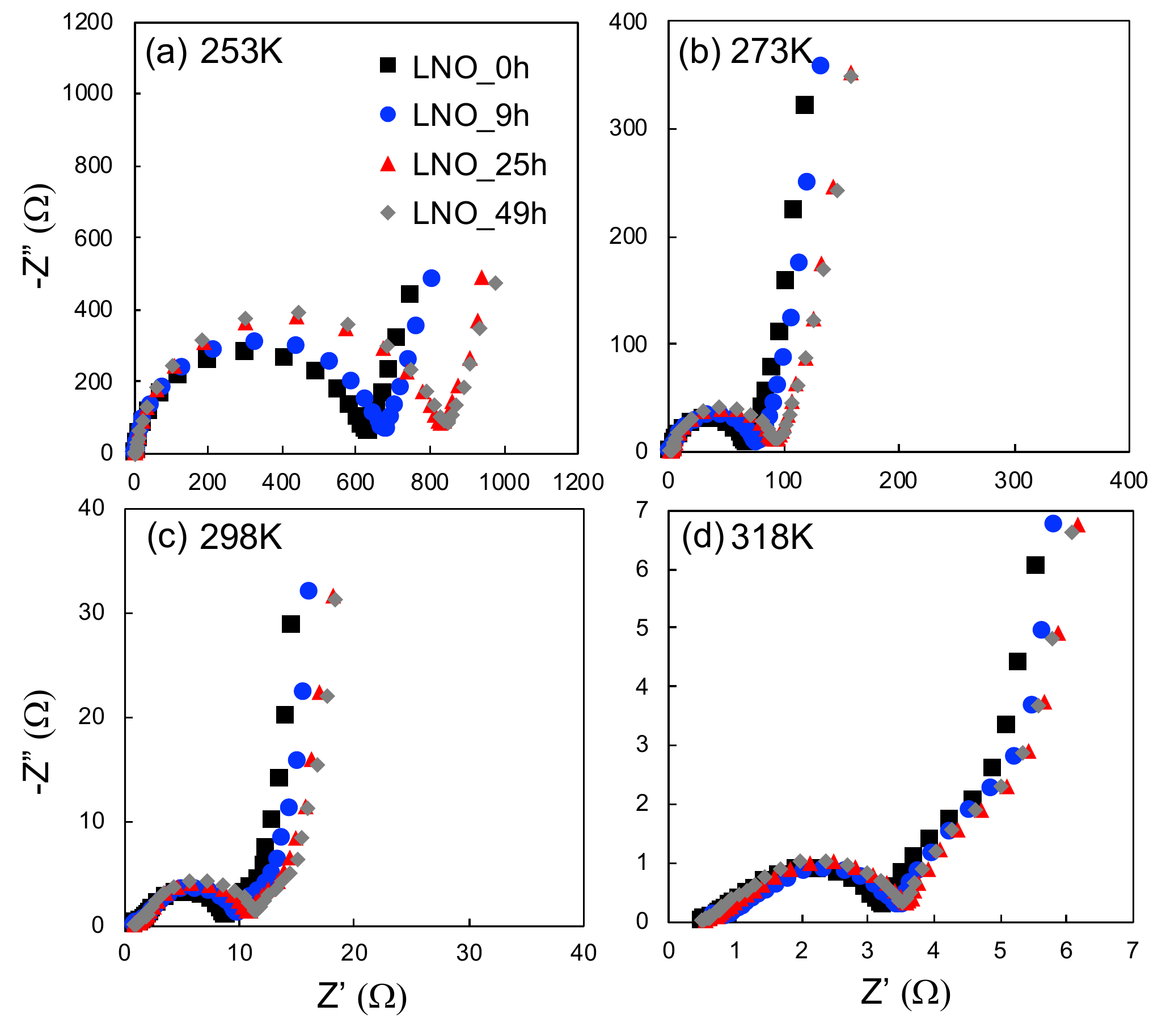}
 \caption{EIS spectrum of \ce{Li$_{0.75}$NiO2}. 
 Each panel shows the results of (a) 253 K, (b) 273 K, (c) 298 K, (d) 318 K}
 \label{fig.fig7}
\end{figure}

EIS measurements were carried out to investigate the electrochemical resistance.
In this study, the positive electrode resistance was measured using a symmetric coin cell with the positive electrodes facing each other.
The resistance components of the cathode and anode are intermixed in the impedance spectrum of a normal full cell, 
and distinguishing the resistance of the positive electrode alone is difficult.
Furthermore, in order to evaluate the relationship between the amount of \ce{Li} contained in \ce{LiNiO2} and the resistance,
measurements were performed under the conditions of \ce{Li$_{0.50}$NiO2} and \ce{Li$_{0.75}$NiO2}.
In addition, EIS measurements were performed under four conditions of 253, 273, 298, and 318 K 
to obtain the temperature dependence of \ce{Li+} diffusion coefficient and the activation energy of the diffusion.
Figure \ref{fig.fig6} and Figure \ref{fig.fig7} show the Nyquist plots of \ce{Li$_{0.50}$NiO2} and \ce{Li$_{0.75}$NiO2}, respectively.
In \ce{Li$_{0.50}$NiO2}, the charge transfer resistance (R$_{\rm ct}$) increased with increase in the amount of \ce{Li+}/\ce{H+} cation exchange in the temperature range of 253--318 K.
On the other hand, in \ce{Li$ _{0.75}$NiO2}, R$_{\rm ct}$ markedly increased in the LNO$\_$25 h and LNO$\_$49 h samples 
in the low-temperature region of 253 K, but no significant difference in R$_{\rm ct}$ was observed in any sample in the region of 273--318 K.
To summarize these results, the higher the \ce{Li} concentration in \ce{LiNiO2} or the lower the operating temperature, 
the greater the effect on Li insertion reaction due to \ce{Li+}/\ce{H+} cation exchange.
Some studies have shown that 
the increase in resistance in the low-temperature range is caused by a decrease in the diffusion rate of \ce{Li+}~\cite{Tan2015,Keefe2019,Jin2017a} 
and the relationship between the \ce{Li} concentration in the bulk and the diffusion coefficient of \ce{Li+}.~\cite{Ven2000,Montoro2004}
Therefore, we believe that the \ce{Li_{1-x}H_xNiO2} phase that underwent structural change due to atmospheric exposure affected the \ce{Li+} diffusion.


\begin{figure}
  \includegraphics[width=3.33in]{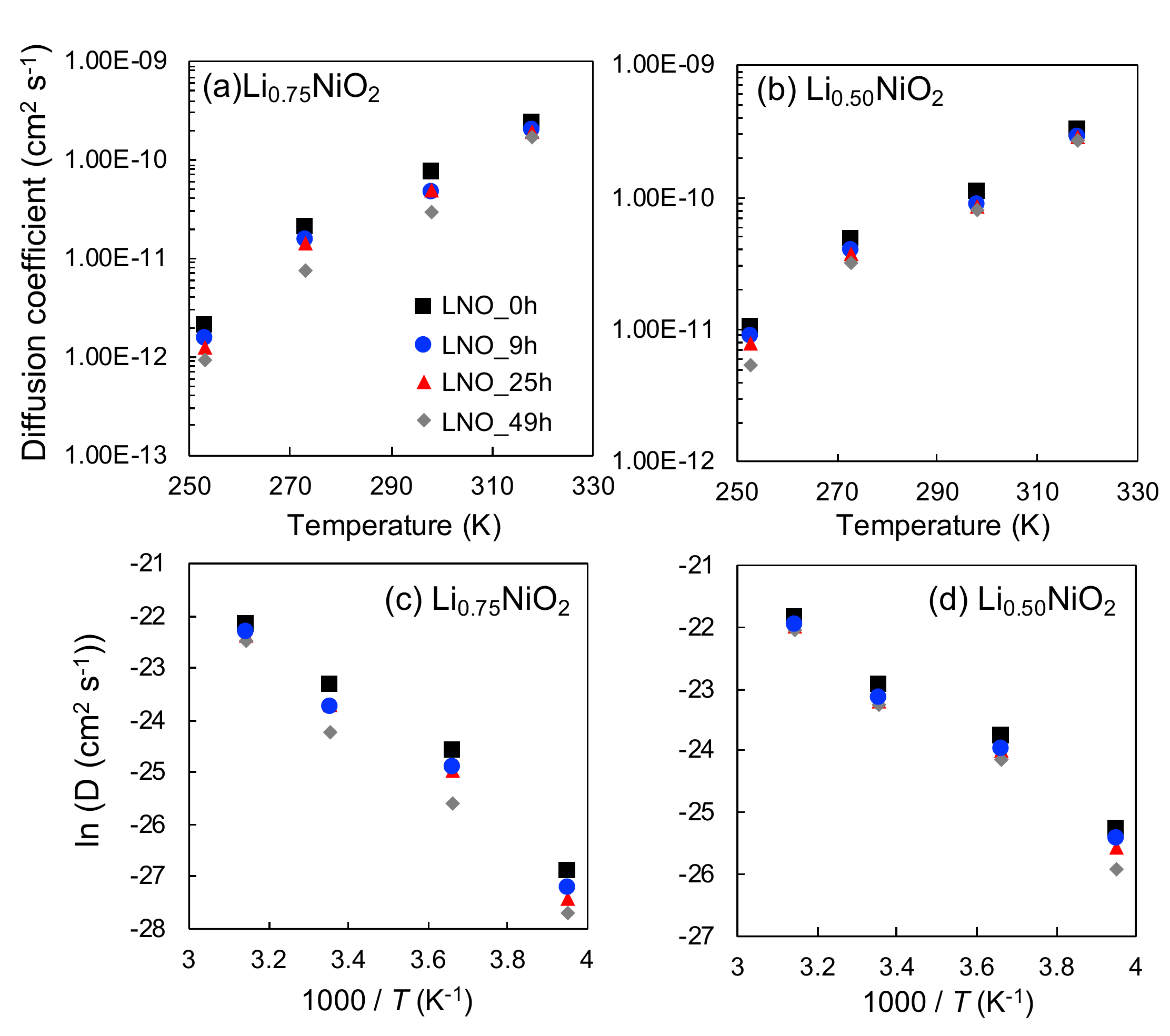}
  \caption{ \ce{Li+} diffusion coefficient determined from EIS measurement. 
               The temperature dependences of $D_{\rm Li^+}$  in \ce{Li$_{0.75}$NiO2} and \ce{Li$_{0.50}$NiO2} are shown in (a) and (b), respectively.
                Panels (c) and (d) show the Arrhenius plots of $D_{\rm Li^+}$ for \ce{Li$_{0.75}$NiO2} and \ce{Li$_{0.50}$NiO2}, respectively.}
  \label{fig.fig8}
\end{figure}

The diffusion region of the EIS spectrum was analyzed to obtain the \ce{Li} diffusion coefficient of each sample.
The linear behavior in the low-frequency region of the EIS spectrum shows the \ce{Li+} diffusion behavior, 
and the \ce{Li+} diffusion coefficient $D_{\rm Li^+}$ is calculated using the expression
eq.(\ref{eq:Li Diffusion}).~\cite{Ho1980,Liu2006a,Chen2016,Jin2017a,Shi2018}
 \begin{equation}
   D_{\rm Li^+} = \frac{R^2 T^2}{2 A^{2}n^{4}F^{4}C^{2}\sigma^{2}},
  \label{eq:Li Diffusion}
 \end{equation}
where $R$ denotes the gas constant, $T$ is the absolute temperature, 
$A$ is the area of the cathode/electrolyte interface ($A$ = 1.33cm$^2$), $n$ is the charge number of the electroactive species (n=1),
$F$ is Faraday's constant, $C$ is the concentration of lithium ion,  and $\sigma$ is the Warburg factor.
The Warburg factor was obtained from the slope of the plots of Z' vs $\omega^{-1/2}$ ($\omega$ is the angular frequency) 
in the Warburg region~.\cite{Liu2006a}
$D_{\rm Li^+}$ obtained by the formula (\ref{eq:Li Diffusion}) is shown in Figure \ref{fig.fig8}.
$D_{\rm Li^+}$ increased with rising temperature, and the calculated value of $D_{\rm Li^+}$ was in good agreement with
the previously reported \ce{LiNiO2} measurement result.~\cite{Nakamura2000,Sugiyama2010a}

Furthermore, we derived the activation energy of diffusion using the following Arrhenius equation.~\cite{Nakamura2000,Okubo2009a,Moriwake2013,Jin2017a}
 \begin{equation}
   D_{\rm Li^+} = D_{\rm 0} \exp \left(\frac{E_{\rm a}}{RT}\right)
 \label{eq:Arrhenius}
 \end{equation}
The determined activation energy ($E_a$) is shown in Table\ref{table2}.
Increase in $E_a$ was confirmed by increasing \ce{Li} concentration in \ce{LiNiO2} and decreasing temperature,
and this result corresponded to the EIS spectrum variation.
Therefore, the influence of \ce{Li+}/\ce{H+} cation exchange on the electrochemical properties (increased resistance and degradation of rate capability) is attributed to the increase in the activation energy of \ce{Li+} diffusion.

\begin{table}
 \caption{Relationship between activation energy and exposure time (\ce{Li+}/\ce{H+} exchange amount) in \ce{Li$_{1-x}$NiO2} (x = 0.25,0.50)}
 \begin{tabular}{lcccc} 
   \hline
             & \multicolumn{4}{c}{$E_a$ (eV)} \\
   \cline{2-5}
   Sample & LNO$\_$0h & LNO$\_$9h & LNO$\_$25h & LNO$\_$49h \\
   \hline
   \ce{Li$_{0.75}$NiO2} & 0.488 & 0.498 & 0.518 & 0.534\\
   \ce{Li$_{0.50}$NiO2} & 0.351 & 0.351 & 0.364 & 0.396\\
   \hline
  \end{tabular}
 \label{table2}
\end{table}

\begin{figure}
 \includegraphics[width=3.33in]{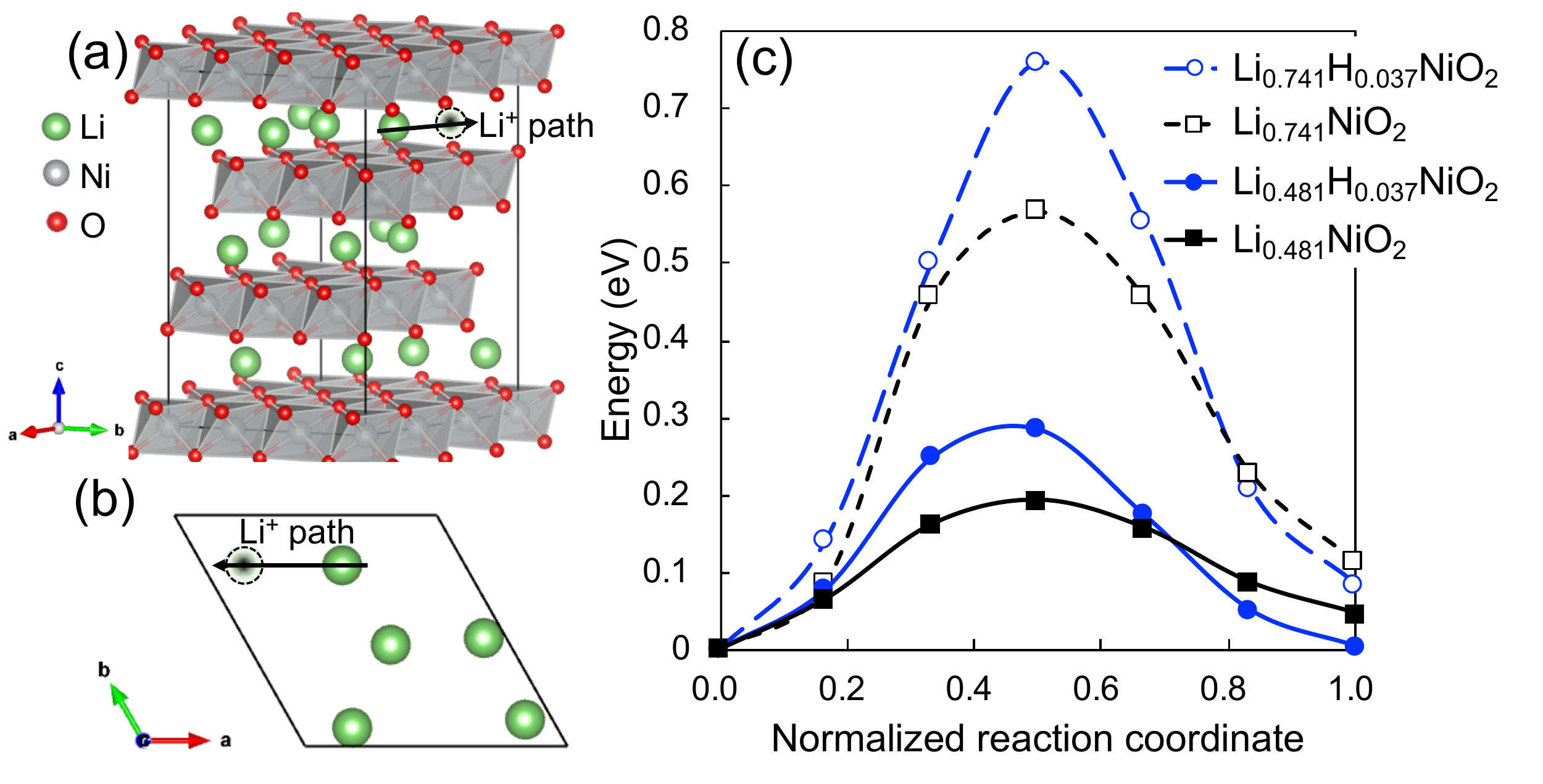}
 \caption{(a): \ce{Li+} diffusion path of \ce{Li$_{0.481}$NiO2} by CI-NEB calculation, 
               (b): (a) Top view of the Li layer, where the Li diffusion path exists in the $c$-axis direction
               (c) Energy profiles of \ce{Li} diffusion path determined by CI-NEB.}
 \label{fig.fig9}
\end{figure}

To further verify whether the change in activation energy was due to the effect of \ce{Li+}/\ce{H+} cation exchange,
the barrier energy for \ce{Li+} diffusion in \ce{Li_{1-x}H_xNiO2} was calculated by the CI-NEB method.
The calculation was performed using a structure in which one of the 27 Li atoms contained 
in the supercell was replaced with an H atom (x = 0.037).
In addition, the barrier energies of \ce{Li$_{0.481}$H_${0.037}$NiO2} and \ce{Li$_{0.741}$H_${0.037}$NiO2}, 
which are close to the amount of Li desorption in the experiment, were calculated.
Figure \ref{fig.fig9}(a)(b) shows the \ce{Li+} diffusion paths of \ce{Li$_{0.481}$NiO2} as an example of the calculated structure.
The energy profiles of \ce{Li$_{0.741}$NiO2}, \ce{Li$_{0.741}$H_${0.037}$NiO2}, 
\ce{Li$_{0.481}$NiO2}, and \ce{Li$_{0.481}$H_${0.037}$NiO2} are shown in Figure \ref{fig.fig9}(c); 
the barrier energies were determined to be 0.567, 0.760, 0.193, and 0.286 eV, respectively.
At the Li concentrations of (1-x) = 0.48 and 0.75, the barrier energy increased with
increase in the \ce{H+} concentration in \ce{LiNiO2}, and the same tendency as observed in the experimental results was obtained.
In literature~\cite{Kang2006}, the barrier energy of \ce{Li$_{0.50}$NiO2} is reported as approximately 0.20--0.30 eV,
and the calculation result of \ce{Li$_{0.481}$NiO2} shows a close value (0.193 eV).

\begin{figure}
 \includegraphics[width=3.33in]{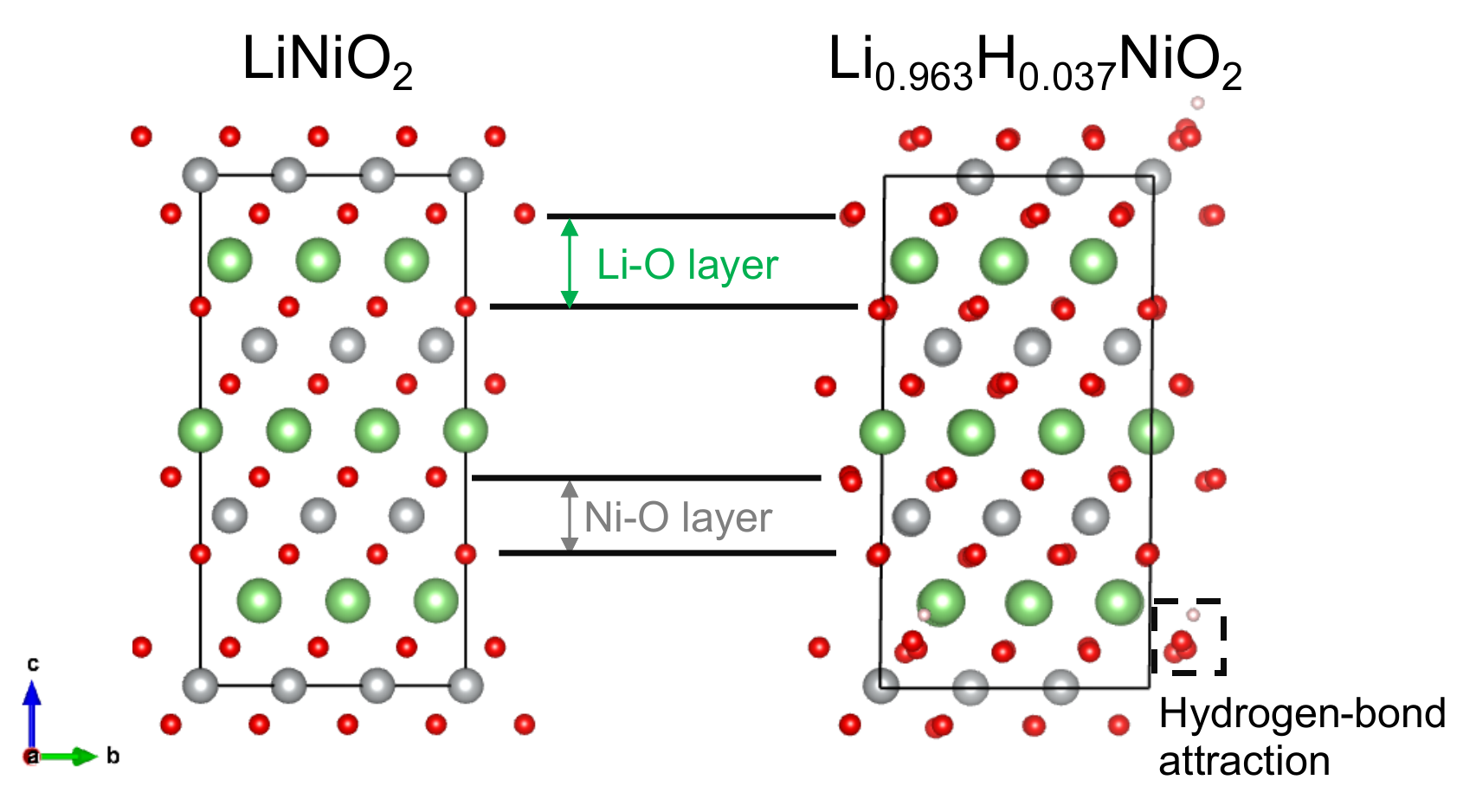}
 \caption{Crystal structure of \ce{LiNiO2} and \ce{Li_${0.963}$H_${0.037}$NiO2} after structural relaxation.}
 \label{fig.fig10}
\end{figure}

After structural relaxation calculation, Figure \ref{fig.fig10} shows that \ce{O} atoms are attracted to \ce{H} atom in the Li-O layer of \ce{Li_${0.963}$H_${0.037}$NiO2}.
In \ce{NiOOH}, it has been reported~\cite{Horanyi1989,VanderVen2006,Li2014,Tkalych2015} that the \ce{NiO2} and \ce{Ni(OH)2} slabs are alternately stacked by bonding through hydrogen bonds.
From the reported phenomenon, it is presumed that the \ce{H} atoms in the Li-O layer cause an attractive action
in the form of hydrogen bonds between the upper and lower \ce{NiO2} slabs, resulting in the shrinkage of the Li-O layer thickness.
In addition, the inverse relationship between the Li-O layer thickness and the activation energy is known, ~\cite{Kang2006}
which is presumed to be the cause of increase in the activation energy due to \ce{Li+}/\ce{H+} cation exchange.

\begin{figure}
 \includegraphics[width=3.33in]{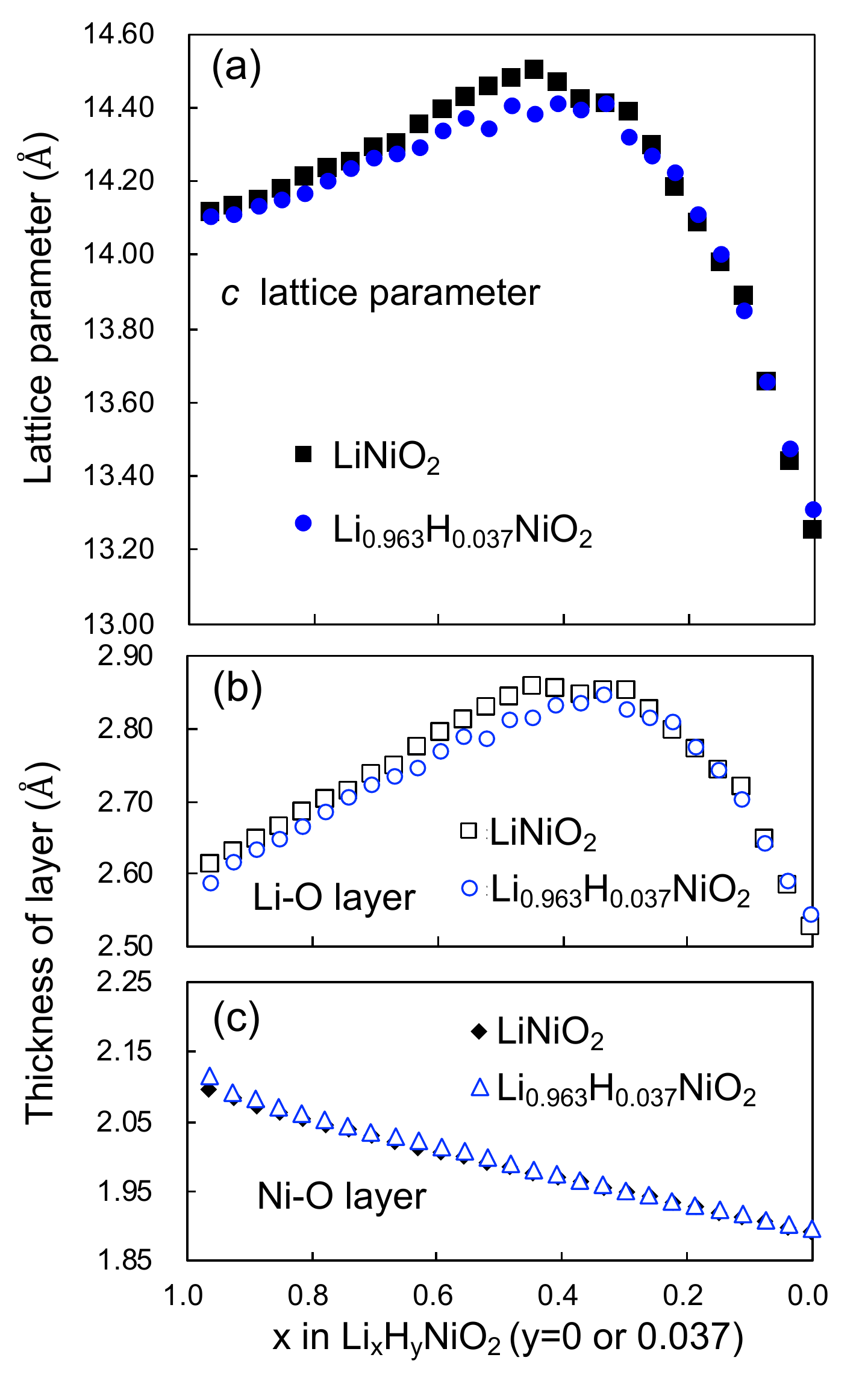}
 \caption{(a): Relationship between \ce{Li} concentration and c-axis lattice constant of \ce{Li_xH_yNiO2}. 
               Panels (b) and (c) show the relationship between the amount of \ce{Li} desorption and the change in thickness of the Li-O and Ni-O layers, respectively.}
 \label{fig.fig11}
\end{figure}

Furthermore, we investigated changes in the c lattice constant, Li-O layer thickness, 
and Ni-O layer thickness with respect to the amount of Li desorption in \ce{Li_xNiO2} and \ce{Li_xH$_{0.037}$NiO2}. 
The structures in which Li was sequentially desorbed one after the other were determined by DFT calculation, 
and the parameters for each structure were obtained.
\ce{Li_xNiO2} has larger $c$ lattice parameter than \ce{Li_xH$_{0.037}$NiO2} 
in the region of x = 0.4--1.0 (Figure \ref{fig.fig11}(a)).
Based on the trend of the change in the layer thicknesses of the Li-O layer and Ni-O layer, 
it was confirmed that the change in the c-axis lattice constant due to \ce{Li+}/\ce{H+} cation exchange was dominated
by the change in thickness of the Li layer (Figure \ref{fig.fig11}(b),(c)).
This result agrees with the change in the activation energy determined by the experiment and CI-NEB method.

\label{section.results and discussion}
\section{SUMMARY}
The structural changes and their effects on the electrochemical properties of \ce{LiNiO2} exposed to the atmosphere were investigated.
Thermal and chemical analyses  revealed that \ce{LiNiO2} exposed to the atmosphere assumes 
the structure of \ce{Li_{1-x}H_xNiO2}  with \ce{Li+}/\ce{H+} cation exchange.
We synthesized \ce{LiNiO2} with different amounts of \ce{H+} exchange, and investigated the influence of \ce{Li_{1-x}H_xNiO2} on the electrochemical properties.
As a result, it was found that the activation energy of \ce{Li+} diffusion increased by the exchange of \ce{H+}, 
resulting in higher resistance and lower rate capability.
Our first principles analysis calculation revealed that the increase in the activation energy was caused 
by decrease in the Li layer thickness, that is, constriction of the diffusion path of \ce{Li+}.
\label{section.sammary}
\section{Acknowledgments}
T. ~T. is deeply grateful to K. ~Ryoshi and T. ~Yoshida for their fruitful discussions and technical support.
The computations in this work were performed 
using the facilities of 
Research Center for Advanced Computing 
Infrastructure at JAIST. 
K.~H. is grateful for the financial support from KAKENHI (grant numbers 17K17762 and 19K05029), 
Grant-in-Aid for Scientific Research on Innovative Areas (16H06439 and 19H05169),
the FLAG-SHIP2020 project (MEXT for computational resources, projects hp190169 and 
hp190167 using K-computer), PRESTO (JPMJPR16NA), and Materials research
by Information Integration Initiative (MI$^2$I) project of the
Support Program for Starting Up Innovation Hub from Japan Science
and Technology Agency (JST).
R.~M. is grateful for financial support from MEXT-KAKENHI (projects JP19H04692 and JP16KK0097), 
the FLAG-SHIP2020 project (MEXT for computational resources, projects hp190169 and 
hp190167 using K-computer),
and the Air Force Office of Scientific Research (AFOSR-AOARD/FA2386-17-1-4049;FA2386-19-1-4015).

\bibliography{references}
\end{document}